\RequirePackage{ifpdf}
\ifpdf 
\documentclass[pdftex]{sigma}
\else
\documentclass{sigma}
\fi

\begin{document}

\renewcommand{\PaperNumber}{***}

\FirstPageHeading

\ShortArticleName{Complete sets of harmonic vortices on a plane}

\ArticleName{Complete sets of circular, elliptic and bipolar harmonic vortices on a plane}

\Author{Pablo L. REND\'ON~$^\dag$ and Eugenio LEY-KOO~$^\ddag$}

\AuthorNameForHeading{P.L. Rend\'on and E. Ley-Koo}

\Address{$^\dag$~Grupo de Ac\'ustica y Vibraciones, Centro de Ciencias Aplicadas y Desarrollo Tecnol\'ogico, Universidad Nacional Aut\'onoma de M\'exico, Ciudad Universitaria, AP 70-186, C.P. 04510, M\'exico D. F., M\'exico}
\EmailD{\href{mailto:email@address}{pablo.rendon@ccadet.unam.mx}}

\Address{$^\ddag$~Instituto de F\'{\i}sica, Universidad Nacional Aut\'onoma de M\'exico, Ciudad Universitaria, AP 20-364, C.P. 01000, M\'exico D. F., M\'exico}
\EmailD{\href{mailto:email@address}{eleykoo@fisica.unam.mx}}
 
\ArticleDates{Received August 1, 2014}

\Abstract{
A class of harmonic solutions to the steady Euler equations for incompressible fluids is presented in two dimensions in circular, elliptic and bipolar coordinates. Since the velocity field is solenoidal in this case, it can be written as the curl of a vector potential, which will then satisfy Poisson's equation with vorticity as a source term. In regions with zero vorticity, Poisson's equation reduces to Laplace's equation, and this allows for the construction of harmonic potentials inside and outside circles and ellipses, depending on the coordinate system. The vector potential is normal to the coordinate plane, and is proportional to the scalar harmonic functions on the plane, thereby guaranteeing that the velocity field is also harmonic and is located on the coordinate plane. The components of the velocity field normal to either a circle or an ellipse are continuous, but the tangential components are discontinuous, so that, in effect, a vortex sheet is introduced at these boundaries. This discontinuity is a measure of the vorticity, normal to the plane and distributed harmonically along the perimeter of the respective circles or ellipses. An analytic expression for the streamlines is obtained which makes visualisation of vortices of various geometries and harmonicities possible. This approach also permits a reformulation of the notion of multipolarity of vortices in the traditional sense of a multipolar expansion of the Green's function for Poisson's equation. As an example of the applicability of this formulation to known vortical structures, Rankine vortices of different geometries are expressed in terms of harmonic functions.}



\section{Introduction}
\label{intro}

Multipolar vortices are known to be an important dynamical feature of a variety of two-dimensional and quasi-geostrophic flows~\cite{McWilliams:1984}. They occur regularly in large-scale geophysical, atmospheric and meteorological flows at high Reynolds numbers. The term itself refers to a set of coherent vortex structures where vorticity is distributed according to characteristic geometrical arrangements. Monopoles, for example, will have a single vorticity extremum, while a dipole and tripole will have, respectively, two and three distinct vorticity extrema. Unstable monopolar structures can relatively easily condense into two isolated dipolar structures. However, it has been also observed that it is possible to nonlinearly destabilize an isolated monopolar vortical structure in order to obtain a single tripolar structure with distributed vorticity~\cite{Carton:1989}. This formation of dipolar and tripolar structures has been observed in laboratory experiments~\cite{VanHeijst:1989}. It has also been shown that higher-order multipolar structures may result from strong perturbation of two-dimensional vortical structures~\cite{Carnevale:1994}. These structures, which we may collectively term $n$-polar vortices, typically consist of a central core of vorticity surrounded, or shielded, by $n-1$ satellite vortices with sign opposite to that of the central vortex, so that the total circulation of the structure is approximately zero, and the whole structure rotates in steady fashion.

Morel and Carton~\cite{Morel:1994} have studied the generation and stability of a variety of multipolar solutions to the two-dimensional Euler equations. For this purpose they simply simulate numerically the generation of higher-order multipolar equilibria taking monopolar shielded vortices as their starting point. Although much of the research concerned with multipolar vortices is either of an experimental or a numerical nature, due to their complicated structure, Crowdy~\cite{Crowdy:1999} has provided a class of exact solutions, which consists of a family of finite-area patches of nonzero vorticity, thus permitting some insight into the shapes of these vortical equilibria which is not possible when considering simple point-vortex models. These solutions exhibit the same properties as experimentally observed multipolar vortices, and, interestingly, only allow for interaction between vortices when these overlap. A generalization, also by Crowdy~\cite{Crowdy:2002a}, of this class of solutions is characterized by an annular region of vorticity enclosing a region of irrotational fluid . It is shown that a nonlinear superposition property allows for the construction of multipolar equilibria having vortical regions with more elaborate geometries and topological structure. Further, both classes of solutions presented by Crowdy~\cite{Crowdy:1999, Crowdy:2002a} share the important property of ''invisibility'' in the sense that they do not induce any irrotational velocity field outside the regions where vorticity is nonzero. Thus, the physical detection of these vortices is not a trivial matter, as they cannot be detected through remote measurements of the far-field velocity.

In this paper we consider an alternative family of exact solutions to the two-dimensional Euler equations, which we associate with the notion of multipolarity in the traditional sense of a series expansion of the Green's function for Poisson's equation on the plane. The solutions thus obtained do not satisfy, in general, the criteria established by Crowdy~\cite{Crowdy:1999}  and Saffman~\cite{Saffman:1992} for vortical equilibria. However, they do constitute, by construction, a complete set of solutions, so that the solutions representing multipolar equilibria as described above may be represented in terms of this family of solutions. They also satisfy the laws of vortex motion as originally set out by Helmholtz~\cite{Helmholtz:1858}. These solutions consist of a closed region of circular or elliptic shape where vorticity sources are continuously distributed along the curve at the border of the region, constituting, in effect, a vortex sheet. As with Rankine vortices and Crowdy's classes of solutions, these solutions represent isolated finite-area regions of nonzero vorticity surrounded by irrotational flow. We stress that these multipolar harmonic expansions are valid and unique both inside and outside the region within which the sources of vorticity are distributed, and, crucially, that sources of a given harmonicity produce velocity fields with the same harmonicity. Ley-Koo and G\'ongora have examined similar multipole expansions inside and outside regions in which sources of either the electrostatic or magnetostatic fields have been distributed~\cite{LeyKoo:1988,LeyKoo:1991}. In fact, it is well-known that the relationship between a vortex filament and the velocity field to which it is associated in an incompressible fluid is formally equivalent to the relationship between a current in a conducting wire and the magnetic field to which it gives rise. This equivalence was already noted by Helmholtz~\cite{Helmholtz:1858}, is discussed in detail by Moffatt~\cite{Moffatt:2008}, and it serves as motivation for many of the ideas and methods explored in this paper.

From a mathematical point of view, it is important and convenient to recognize the superintegrability and exact solvability of the Euler equations. In the specific case of vortices in two dimensions on a plane, their solutions involve the familiar circular functions in the angular coordinates, and power functions and exponential or hyperbolic functions in the radial coordinates for the successive geometries. On the other hand, this article is intended for the fluid physics readership, and we have thus chosen to formulate it using the familiar methods of differential equations. Correspondingly, in section~\ref{basics} we include the differential equations relating the solenoidal vector fields of velocity, vorticity, and velocity potential, as well as the integral forms of the appropriate boundary conditions which will permit us to distribute vorticity along circular and elliptic boundaries. In section~\ref{laplace} we write both Laplace's equation in polar, elliptic, and bipolar coordinates, and its associated solutions, and in section~\ref{harmonicvortices} we use these solutions to derive the exact, explicit form of the vorticity and velocity fields associated with harmonic vortices both inside and outside the given boundaries for all three sets of coordinates. Graphical representation of these velocity fields is given in section~\ref{graphical} in the form of families of streamlines illustrating distinctive features of the vortical structures for each coordinate system. Using the Green's functions for Poisson's equation in all three coordinate systems, we express in section~\ref{linevortices} a well-known stable vortex solution of the Euler equations, the Rankine vortex, in terms of our harmonic solutions, highlighting the usefulness of constructing a complete set of harmonic solutions to these equations. Finally, we discuss, in section~\ref{discussion}, connections with other types of two-dimensional vortices on planes and spheres, as well as extensions to three dimensions and other geometries.    

\section{Velocity, vorticity, and vector potential}
\label{basics}

Following Saffman~\cite{Saffman:1992}, we describe the motion of an incompressible fluid through the velocity field, $\vec{u}=\vec{u}(\mathbf{x},t)$, and we define the vorticity $\vec{\omega}=\vec{\omega}(\mathbf{x},t)$ as follows:

\begin{equation}
\vec{\omega} = \nabla \times \vec{u},
\label{circulation}
\end{equation}

\noindent
so that, by construction, the vorticity is solenoidal. Notice that, if we take $\Gamma$ to be the circulation of $\vec{u}$ for the closed curve $C$, we have

\begin{equation}
\Gamma = \int_S \vec{\omega} \cdot \vec{dS} = \oint_C \vec{u} \cdot \vec{d \ell},
\label{vorticityintegral}
\end{equation}

\noindent
where $S$ is the surface element delimited by the closed curve $C$. Thus, we may interpret the vorticity as a circulation density per unit area. Further, in an incompressible fluid, the velocity field is also solenoidal, that is,

\begin{equation}
\nabla \cdot \vec{u} = 0.
\label{flux}
\end{equation}

\noindent
In this case, we may also define a vector potential $\vec{A}$ such that

\begin{equation}
\vec{u} = \nabla \times \vec{A},
\label{vectorpotential}
\end{equation}

\noindent
allowing us to write

\begin{equation}
\vec{\omega} = \nabla \times \left ( \nabla \times \vec{A} \right) = \nabla \left( \nabla \cdot \vec{A} \right) - \nabla^2 \vec{A}.
\label{vectoridentity}
\end{equation}

\noindent
By fixing the gauge of $\vec{A}$ so that $\nabla \cdot \vec{A} = 0$, we observe that equation~(\ref{vectoridentity}) is, in effect, Poisson's equation,

\begin{equation}
\nabla^2 \vec{A} = - \vec{\omega}.
\label{poisson}
\end{equation}

Equations (\ref{flux}) and (\ref{circulation}) can be written in the form of conditions to be satisfied along a given boundary. Equation (\ref{flux}) can be recast in the integral forms,

\begin{equation}
\oint_S \vec{u} \cdot \vec{dS} = \left( \vec{u_2} - \vec{u_1} \right) \cdot \hat{n} \, \Delta S = 0,
\label{fluxbc}
\end{equation}

\noindent
where $\vec{u_1}$ and $\vec{u_2}$ are the velocities on either side of the boundary, enclosed by a Gauss pill-box with areas $\pm \hat{n} \, \Delta S$. This form of the equation clearly expresses the continuity of the normal components of the velocity across the boundary.

Equations (\ref{circulation}) and (\ref{vorticityintegral}), on the other hand, can be used to write:

\begin{equation}
\oint_C \vec{u} \cdot \vec{d \ell} = \left( \vec{u_2} - \vec{u_1} \right) \cdot \hat{t} \, \Delta \ell = \Delta \Gamma,
\label{circulationbc}
\end{equation}

\noindent
for a Stokes rectangular circuit $C$ with tangential elements $\pm \hat{t} \, \Delta \ell$. Thus, the tangential components of the velocity are discontinuous across the boundary, and this discontinuity gives the measure of the circulation per unit length, $\Delta \Gamma / \Delta \ell$.

\section{Laplace's equation in polar, elliptic, and bipolar coordinates}
\label{laplace}

Our objective now is to write the Laplacian operator in terms of the three sets of coordinates described in the Appendix. We shall denote, in general form, the two independent coordinates on the plane as $q_1$, and $q_2$, with $\hat{e}_1$ and $\hat{e}_2$ the unit vectors normal to lines of constant values of the respective coordinates, and $dq_1$ and $dq_2$ their corresponding variations in these directions. In our case, $(\hat{e}_{1},\hat{e}_{2}) = (\hat{r},\hat{\varphi}) = (\hat{u},\hat{v}) = (\hat{\sigma},\hat{\tau})$ for polar, elliptic, and bipolar coordinates, respectively. Note that the unitary vectors that constitute each coordinate pair are always orthogonal. The displacement vector $d\vec{\ell}$ can then be expressed in the following form:

\begin{equation}
d\vec{\ell}=\hat{e}_{1}h_{1}dq_{1}+\hat{e}_{2}h_{2}dq_{2},
\label{displacement}
\end{equation}

\noindent
where $h_1$ and $h_2$ are scale factors, and their values, for the three coordinate systems under consideration, are given below.

\begin{align*}
& \mbox{Polar coordinates:} &&h_1 = 1, \, h_2 = r. \\
& \mbox{Elliptic coordinates:} &&h_1 = h_2 = f \, \sqrt{\cosh^2 u - \cos^2 v}. \\
& \mbox{Bipolar coordinates:} &&h_{1} = h_{2} = a / ( \cosh \tau - \cos \sigma).
\end{align*}

\noindent
The Laplacian operator can now be written in terms of curvilinear orthogonal coordinates as follows:

\begin{equation}
\nabla ^{2}=\frac{1}{h_{1}h_{2}}\left[ \frac{\partial }{\partial q_{1}}\frac{
h_{2}}{h_{1}}\frac{\partial }{\partial q_{1}}+\frac{\partial }{\partial q_{2}
}\frac{h_{1}}{h_{2}}\frac{\partial }{\partial q_{2}}\right],
\label{laplacian}
\end{equation}

\noindent
so that, for the corresponding coordinate systems, Laplace's equation takes the following forms:
 
\begin{align}
\left[ \frac{1}{r}\frac{\partial }{\partial r} r\frac{\partial }{\partial r}+ \frac{1}{r^{2}}\frac{\partial ^{2}}{\partial \varphi ^{2}}\right] \Phi \left( r,\varphi \right) & = 0,  
\label{laplace1} \\
\frac{1}{f^{2}\left( \cosh ^{2}u-\cos ^{2}v\right) }\left[ \frac{\partial ^{2}}{\partial u^{2}}+\frac{\partial ^{2}}{\partial v^{2}}\right] \Phi \left( u,v\right) & = 0,  
\label{laplace2} \\
\frac{\left( \cosh \tau -\cos \sigma \right)^{2}}{a^{2}} \left[ \frac{ \partial ^{2}}{\partial \tau ^{2}}+\frac{\partial ^{2}}{\partial \sigma ^{2}} \right] \Phi \left( \tau ,\sigma \right) & = 0.
\label{laplace3}
\end{align}

\noindent
The solutions to these equations are given in terms of the circular, elliptic, and bipolar harmonics, respectively, as

\begin{align}
\Phi _{m}\left( r,\varphi \right) &= \left( A_{m} r^{m}+B_{m} r^{-m}\right)
\left( 
\begin{array}{c}
\sin m\varphi \\ 
\cos m\varphi
\end{array}
\right), \label{harmonic1} \\
\Phi _{m}\left( u,v\right) &= \left( A_{m}e^{mu}+B_{m}e^{-mu}\right) \left( 
\begin{array}{c}
\sin mv \\ 
\cos mv
\end{array}
\right), \label{harmonic2} \\
\Phi _{m}\left( \tau ,\sigma \right) &= \left( A_{m}e^{m\tau
}+B_{m}e^{-m\tau }\right) \left( 
\begin{array}{c}
\sin m\sigma \\ 
\cos m\sigma
\end{array}
\right), \label{harmonic3}
\end{align}

\noindent
for $m=1,2,3,\ldots$ Isotropic solutions, which occur when $m=0$, correspond to the choice of the cosine -- which takes on the value 1 -- as the angular function, and the radial functions

\begin{align}
\Phi _{0}\left( r \right) &= A_{0}+B_{0}\ln r,  \label{zeroharmonic1} \\
\Phi _{0}\left( u\right) &= \left( A_{0}+B_{0}u\right),  \label{zeroharmonic2} \\
\Phi _{0}\left( \tau \right) &= \left( A_{0}+B_{0}\tau \right), \label{zeroharmonic3}
\end{align}

\noindent
which is logarithmic in the circular case, and linear in the elliptic and bipolar cases.

\section{Harmonic vortices inside and outside coordinate circles and ellipses}
\label{harmonicvortices}

In this section we construct explicitly the velocity and vorticity fields associated with harmonic vortices both inside and outside coordinate circles and ellipses corresponding to the use of circular and bipolar coordinates on the one hand, and elliptic coordinates on the other.

For vortices on a plane, we consider the vector velocity potential $\vec{A}$ to point in the direction of $\hat{k}$, perpendicular to the $x-y$ plane. These functions, as we have seen in section~\ref{basics}, must be harmonic and bounded both inside and outside the corresponding boundaries, circles or ellipses. The velocity can be obtained from this vector velocity potential in a straightforward manner, as follows:

\begin{equation}
\vec{u}\left( q_{1},q_{2}\right) =\nabla \times \vec{A}=\frac{1}{h_{1}h_{2}}
\begin{vmatrix}
h_{1}\hat{e}_{1} & h_{2}\hat{e}_{2} & \hat{k} \\ 
\frac{\partial }{\partial q_{1}} & \frac{\partial }{\partial q_{2}} & \frac{
\partial }{\partial z} \\ 
0 & 0 & A_{z}
\end{vmatrix}.
\label{curl}
\end{equation}

\noindent
Notice that in this formulation the vector potential formally incorporates the role normally assigned to the stream function. Finally, once we have the velocity, we can plot the corresponding streamlines to visualize the flow around the vortices. The equations for the streamlines are arrived at by writing the velocity as

\begin{equation}
\vec{u}=\hat{e}_{1}u_{1}+\hat{e}_{2}u_{2;},
\label{velocity}
\end{equation}

\noindent
and observing that the components of $\vec{u}$ must be proportional to the respective components of $d\vec{\ell}$, in the form of (\ref{displacement}), at every point. Thus, we obtain

\begin{equation}
\frac{h_{1}dq_{1}}{u_{1}}=\frac{h_{2}dq_{2}}{u_{2}}.
\label{streamlines}
\end{equation}

We will now proceed to write the vector velocity potential for each of the three coordinate systems, and to obtain the velocity and the streamline equations, both inside and outside the appropriate borders. Superindices $i$ and $e$ are used to denote whether the velocity potential and velocity correspond to the interior or exterior regions, respectively.

\subsection{Vortices in polar coordinates}
\label{vortpolar}

When $m=0$ we set

\begin{align}
\vec{A}_{0}^{i} &= \hat{k}A_{0}^{i},
\label{polari0} \\
\vec{A}_{0}^{e} &= \hat{k} A_{0}^{e} \ln r,  
\label{polare0}
\end{align}

\noindent
and when $m \geqslant 1$ we set

\begin{align}
\vec{A}_{m}^{i} &= \hat{k}A_{m}^{i}r^{m}\left( 
\begin{array}{c}
\cos m\varphi \\ 
\sin m\varphi
\end{array}
\right),
\label{polarim} \\
\vec{A}_{m}^{e} &= \hat{k}A_{m}^{e}r^{-m}\left( 
\begin{array}{c}
\cos m\varphi \\ 
\sin m\varphi
\end{array}
\right).
\label{polarem}
\end{align}

\noindent
Since we require the vector potential to be continuous at the boundary, using equations (\ref{polari0})-(\ref{polarem}) we impose the following conditions at $r=r_0$:

\begin{align}
A_{0}^{i} & = A_{0}^{e} \ln r_{0}=A_{0} \ln r_{0},
\label{bcpolar1} \\
A_{m}^{i} r_{0}^{m} & = A_{m}^{e} r_{0}^{-m}=A_{0},
\label{bcpolar2}
\end{align}

\noindent
where $A_0$ is a constant. Equations (\ref{polari0})-(\ref{polarem}) may now be written in a more compact form, as follows:

\begin{align}
\vec{A}_{0}^{i,e} & = \hat{k} A_{0} \ln r_>,
\label{polar0} \\
\vec{A}_{m}^{i,e} & = \hat{k}A_0 \frac{r_{<}^m}{r_{>}^m} \left( 
\begin{array}{c}
\cos m\varphi \\ 
\sin m\varphi
\end{array}
\right),
\label{polarm}
\end{align}

\noindent
where $r_{<} = \min \{ r, r_0 \}$ and $r_{>} = \max \{ r, r_0 \}$.

The corresponding velocity fields are then obtained by means of equation~(\ref{curl}) and are given, for $m=0$, by

\begin{align}
\vec{u}^{\, i}_0 & = \vec{0},
\label{polarvelocityi0} \\
\vec{u}^{\, e}_0 & = - \hat{\varphi} \frac{A_0}{r}, 
\label{polarvelocitye0}
\end{align}

\noindent
and, for $m \geqslant 1$, by

\begin{align}
\vec{u}_m^{\, i} & = A_0 \, m \frac{r^{m-1}}{r_0^m} \left( 
\begin{array}{c}
- \hat{r} \sin m \varphi - \hat{\varphi} \cos m\varphi \\ 
\hat{r} \cos m \varphi - \hat{\varphi} \sin m\varphi 
\end{array}
\right),
\label{polarvelocityim} \\
\vec{u}_m^{\, e} & = A_0 \, m \frac{r_0^m}{r^{m+1}} \left( 
\begin{array}{c}
- \hat{r} \sin m \varphi + \hat{\varphi} \cos m\varphi \\ 
\hat{r} \cos m \varphi + \hat{\varphi} \sin m\varphi 
\end{array}
\right).
\label{polarvelocityem}
\end{align}

\noindent
Note that while the radial components are continuous at the circular boundary defined by $r=r_0$, the tangential components are discontinuous.

Using equation~(\ref{streamlines}) we obtain the analytical form of the streamlines inside the circular boundary. When $m=0$, we see from equation~(\ref{polarvelocityi0}) that inside the circle there are no streamlines, and from equation~(\ref{polarvelocitye0}) that outside the circle the streamlines are circles with radius $r>r_0$. When $m \geqslant 1$, the streamlines are obtained from equations (\ref{polarvelocityim}) and (\ref{polarvelocityem}), according to the orientation of the velocity field, and are given either by

\begin{align}
r^m \cos m \varphi & = r_0^m \cos m \varphi_0  &\mbox{for } r<r_0,
\label{polarstreamin1} \\
r^{-m} \, \cos m \varphi & =  r_0^{-m} \, \cos m \varphi_0 &\mbox{for } r>r_0,
\label{polarstreamout1} 
\end{align}

\noindent
or by

\begin{align}
r^m \sin m \varphi & = r_0^m \sin m \varphi_0 &\mbox{for } r<r_0,
\label{polarstreamin2} \\
r^{-m} \, \sin m \varphi & = r_0^{-m} \, \sin m \varphi_0 &\mbox{for } r>r_0,
\label{polarstreamout2} 
\end{align}

\noindent
where $\varphi_0$ is an arbitrary angle. As we can see from the equations above, the dependence of the angular part of the velocities $\vec{u}_m^{\, i,e}$ on either $\cos m \varphi$ or $\sin m \varphi$ determines the harmonicity of each family of streamlines, which is inherited directly from the harmonicity of the vorticity.

\subsection{Vortices in elliptic coordinates}
\label{vortelliptical}

The same criteria applied in the preceding section are used here to arrive at the expressions for the vector velocity potential both inside and outside the boundaries, which in this case are ellipses with major axes aligned along the $x$ axis. Using the same notation as before, with $u_{<} = \min \{ u, u_0 \}$ and $u_{>} = \max \{ u, u_0 \}$ for a given $u_0$, these potentials are given by

\begin{align}
\vec{A}_{0}^{i,e} & = \hat{k} A_{0} \, u_>,
\label{elliptical0} \\
\vec{A}_{m}^{i,e} & = \hat{k}A_0 \, e^{-m u_>} \left( 
\begin{array}{c}
\cosh m u_< \cos mv \\ 
\sinh m u_< \sin mv
\end{array}
\right).
\label{ellipticalm}
\end{align}

The corresponding velocity fields are obtained as before, and they are given, for $m=0$, by

\begin{align}
\vec{u}^{\, i}_0 & = \vec{0},
\label{ellipticalvelocityi0} \\
\vec{u}^{\, e}_0 & = -\frac{\hat{v} \, A_{0}}{f\sqrt{\cosh ^{2}u-\cos ^{2}v}}, 
\label{ellipticalvelocitye0}
\end{align}

\noindent
and, for $m \geqslant 1$, by

\begin{align}
\vec{u}_m^{\, i} & = A_0 m \left( 
\begin{array}{c}
- \hat{u} \cosh mu \sin mv - \hat{v} \sinh mu \cos mv \\ 
\hat{u} \sinh mu \cos mv - \hat{v} \cosh mu \sin mv 
\end{array}
\right) \notag \\
& \times \frac{e^{-m u_0}}{h_u}, 
\label{ellipticalvelocityim} \\
\vec{u}_m^{\, e} & = A_0 m \left( 
\begin{array}{c}
\cosh mu_0 \, ( - \hat{u}  \sin mv + \hat{v} \cos mv )\\ 
\sinh mu_0 \, ( \hat{u} \cos mv + \hat{v} \sin mv
\end{array}
\right) \notag \\
& \times \frac{e^{-m u}}{h_u}, 
\label{ellipticalvelocityem}
\end{align}

\noindent
where $h_u=f\sqrt{\cosh ^2 u-\cos ^2 v}$. Note again that the normal components are continuous at the boundary while the tangential components are not.

Once again, when $m=0$ there are no streamlines inside the elliptic boundary, and the streamlines outside the boundary, which corresponds to $u=u_0$, are confocal coordinate ellipses. When $m \geqslant 1$, we obtain from (\ref{ellipticalvelocityim}) and (\ref{ellipticalvelocityem}) the analytic expression for the streamlines, given either by

\begin{align}
\cosh mu \cos mv & = \cosh mu_0 \cos mv_0 &\mbox{for } u<u_0,
\label{ellipticalstreamin1} \\
e^{-mu} \cos mv & = e^{-mu_0} \cos mv_0 &\mbox{for } u>u_0,
\label{ellipticalstreamout1} 
\end{align}

\noindent
or by

\begin{align}
\sinh mu \sin mv & = \sinh mu_0 \sin mv_0 &\mbox{for } u<u_0,
\label{ellipticalstreamin2} \\
e^{-mu} \sin mv & = e^{-mu_0} \sin mv_0 &\mbox{for } u>u_0,
\label{ellipticalstreamout2} 
\end{align}

\noindent
where $v_0$ is a constant. Again, the distribution of the streamlines depends on whether the tangential part of the velocity is of the form $\cos mv$ or $\sin mv$.

\subsection{Vortices in bipolar coordinates}
\label{vortbipolar}

While the boundaries for both polar and elliptic coordinates consisted solely of one closed curve, in bipolar coordinates the boundary consists of two circles, which correspond to $\tau = \pm \tau_0$, where $\tau_0$ is a positive constant. Note that when $\tau=0$, the boundary coincides with the $y$ axis. The two circles are located symmetrically one on each side of the $y$ axis, and the regions interior and exterior to the circles are given by $| \tau | > | \tau_0 |$ and $| \tau | < | \tau_0 |$, respectively. Taking this into account, if $|\tau| > |\tau_0|$ we define $\tau_{<} = \tau_0$ and $\tau_{>} = \tau$, and, conversely, if $|\tau| < |\tau_0|$ we define $\tau_{<} = \tau$ and $\tau_{>} = \tau_0$. It is now possible to write the vector velocity potentials as

\begin{align}
\vec{A}_{0}^{i,e} & = \hat{k} A_{0} \, \tau_<,
\label{bipolar0} \\
\vec{A}_{m}^{i,e} & = \hat{k}A_0 \, \sinh m \tau_< \, e^{- \mbox{\footnotesize sgn}(\tau_>) m \tau_>} \left( 
\begin{array}{c}
\cos m \sigma \\ 
\sin m \sigma
\end{array}
\right).
\label{bipolarm}
\end{align}

The corresponding velocity fields are, for $m=0$,

\begin{align}
\vec{u}^{\, i}_0 & = \vec{0},
\label{bipolarvelocityi0} \\
\vec{u}^{\, e}_0 & = -\hat{\sigma} \frac{A_0}{a} \left( \cosh \tau - \cos \sigma \right), 
\label{bipolarvelocitye0}
\end{align}

\noindent
and, for $m \geqslant 1$,

\begin{align}
\vec{u}_m^{\, e} & = A_0 m \sinh m \tau_0 \left( 
\begin{array}{c}
- \hat{\tau} \sin m \sigma + \hat{\sigma} \, \mbox{\small sgn} ( \tau ) \, \cos m \sigma \\ 
\hat{\tau} \cos m \sigma + \hat{\sigma} \, \mbox{\small sgn} ( \tau ) \, \sin m \sigma
\end{array}
\right) \notag \\
& \times \frac{e^{- \mbox{\footnotesize sgn} ( \tau ) \, m \tau}}{h_{\tau}}, 
\label{bipolarvelocityim} \\
\vec{u}_m^{\, i} & = A_0 m \left( 
\begin{array}{c}
- \hat{\tau} \sinh m \tau \sin m \sigma - \hat{\sigma} \cosh m \tau \cos m \sigma \\ 
\hat{\tau} \sinh m \tau \cos m \sigma - \hat{\sigma} \cosh m \tau \sin m \sigma 
\end{array}
\right) \notag \\
& \times \frac{e^{- \mbox{\footnotesize sgn} ( \tau_0 ) \, m \tau_0}}{h_{\tau}},
\label{bipolarvelocityem}
\end{align}

\noindent
whre $h_{\tau} = a / ( \cosh \tau - \cos \sigma )$. As expected, the normal components of the velocities are continuous across the boundary, $\tau = \tau_0$, while the tangential components are discontinuous.

As before, when $m=0$, there are no streamlines inside the boundary, which in this case corresponds to the regions inside the two circles. On the outside of the circles, the streamlines are nested coordinate circles corresponding to different values of $\tau < \tau_0$. When $m \geqslant 1$, the families of streamlines are given either by

\begin{align}
e^{\, - \mbox{\footnotesize sgn}( \tau ) m \tau } \, \cos m \sigma & = e^{\, - \mbox{\footnotesize sgn}( \tau_0 ) m \tau_0 } \, \cos m \sigma_0 & \mbox{for } | \tau | > | \tau_0 |,
\label{bipolarstreamin1} \\
\sinh m \tau \, \cos m \sigma & = \sinh m \tau_0 \, \cos m \sigma_0  &\mbox{for } | \tau | < | \tau_0 |,
\label{bipolarstreamout1} 
\end{align}

\noindent
or by

\begin{align}
e^{- \mbox{\footnotesize sgn}( \tau ) m \tau } \, \sin m \sigma & = e^{- \mbox{\footnotesize sgn}( \tau_0 ) m \tau_0 } \, \sin m \sigma_0 & \mbox{for } | \tau | > | \tau_0 |,
\label{bipolarstreamin2} \\
\sinh m \tau \sin m \sigma & =  \sinh m \tau_0 \sin m \sigma_0 &\mbox{for } | \tau | < | \tau_0 |,
\label{bipolarstreamout2} 
\end{align}

\noindent
where $\sigma_0$ is an arbitrary angle. These equations represent, once more, families of streamlines with different harmonicities depending on the form taken by the tangential part of the velocities.

\subsection{Harmonicity and distribution of vorticity}

An important property derived from the analysis above is that, regardless of the choice of coordinate system, the harmonicity of the initial vector potential is inherited by the velocity field, and is in turn reflected in the form taken by the equations describing the streamlines. The sets of solutions for the velocity field, both inside and outside the boundaries, constitute complete harmonic bases which allow for the representation of any number of vortical structures in the form of multipolar expansions.  

In the previous sections, the tangential components of the velocity are shown to be discontinuous at the boundaries, which are either circles or ellipses. Thus, the surfaces formed by the totality of the vortex filaments that pass through these boundaries are, in fact, vortex sheets. These discontinuities represent, according to equation~(\ref{circulationbc}), a measure of the distribution along the boundaries of circulation per unit length, and thus, of vorticity itself. Indeed, we may write the distributed vorticity for all of the above coordinate systems as

\begin{equation}
\vec{\omega} (q_1, q_2) = \hat{k} \, \frac{\delta ( q_1 - q_{10} ) \, \Gamma (q_{10},q_2)}{h_1 \, h_2},
\label{generalvorticity}
\end{equation}

\noindent
where $q_1$ is the coordinate that varies in the direction normal to the boundary, $q_{10}$ is the value of this coordinate on the boundary, and $\Gamma(q_{10},q_2)$ is the intensity of the vortex filament which passes through the point on the boundary at which $q_1 = q_{10}$. It should be stressed that the boundaries do not coincide with streamlines of the flow, as is often the case, but, rather, they represent regions of non-zero vorticity. Indeed, the vorticity is distributed solely along the boundaries, and as a result the velocity field is nonsingular in the whole of both the interior and the exterior regions.

\begin{figure}[!htb]
\centering
\includegraphics[width=9cm]{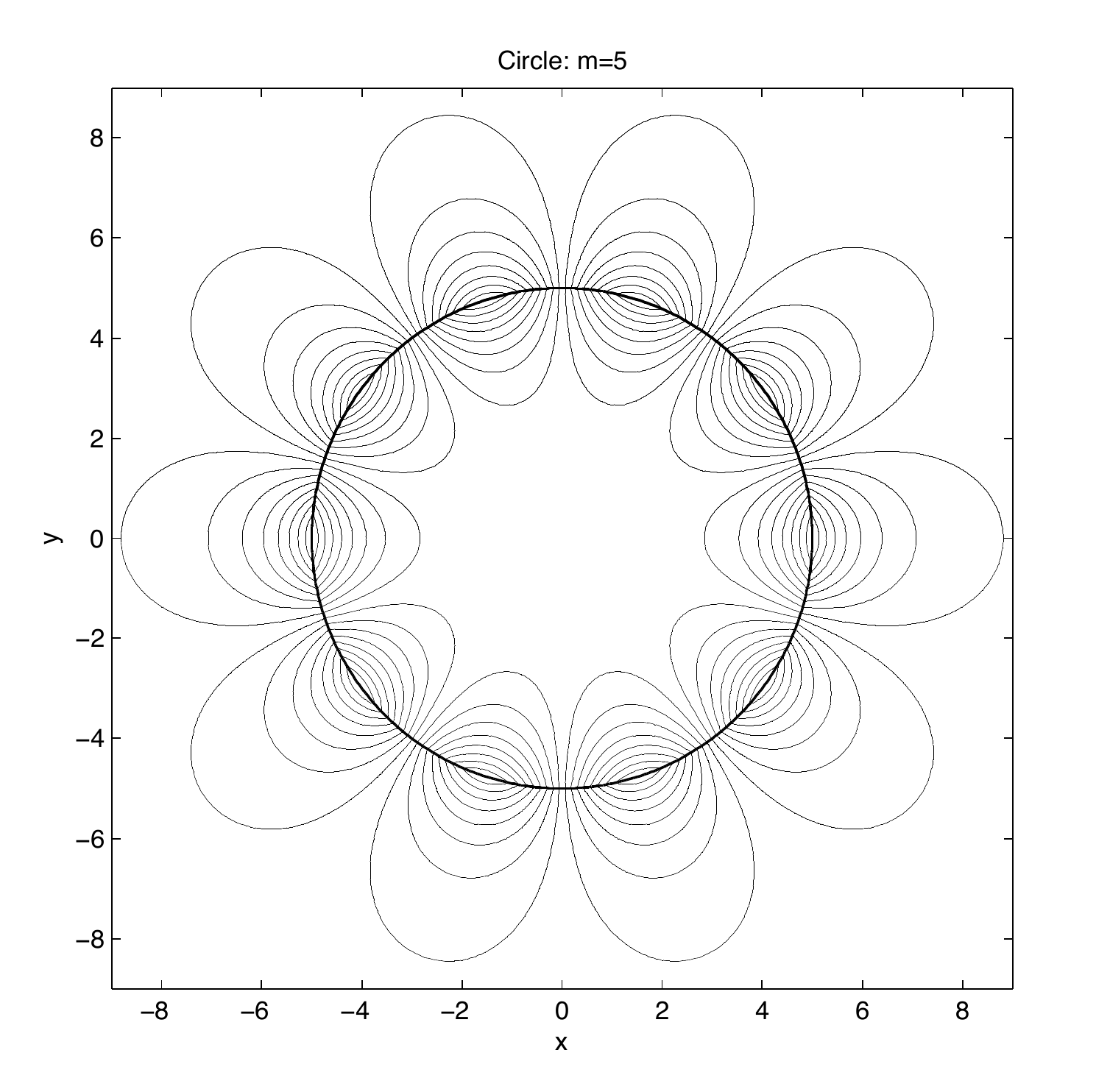}
\caption{\label{polar5} Typical streamlines corresponding to multipolar vortices around a circle, with $m=5$. The harmonicity of the associated velocity field is given by $\cos m \varphi$.}
\end{figure}

\section{Graphical representation of the solutions}
\label{graphical}

In this section we shall plot families of streamlines illustrating different features of the vortical structures for each coordinate system. In all cases we plot the boundaries, which are either ellipses or circles, and a number of streamlines both inside and outside said boundaries. It may be observed, in general, that streamlines arriving at the boundary at very oblique angles experience a marked change in direction as they go across the boundary, whereas streamlines arriving at the boundary at almost the normal direction to the boundary experience very small changes of direction as they go across the boundary. This is completely in accordance with the results obtained in the previous section regarding the continuity of the normal components of the velocities at the boundary, and the discontinuity of the tangential components of the same velocities at the boundary. The boundary circles and ellipses essentially constitute continuous sources of vorticity.

\begin{figure}[!hbt]
\centering
\includegraphics[width=9cm]{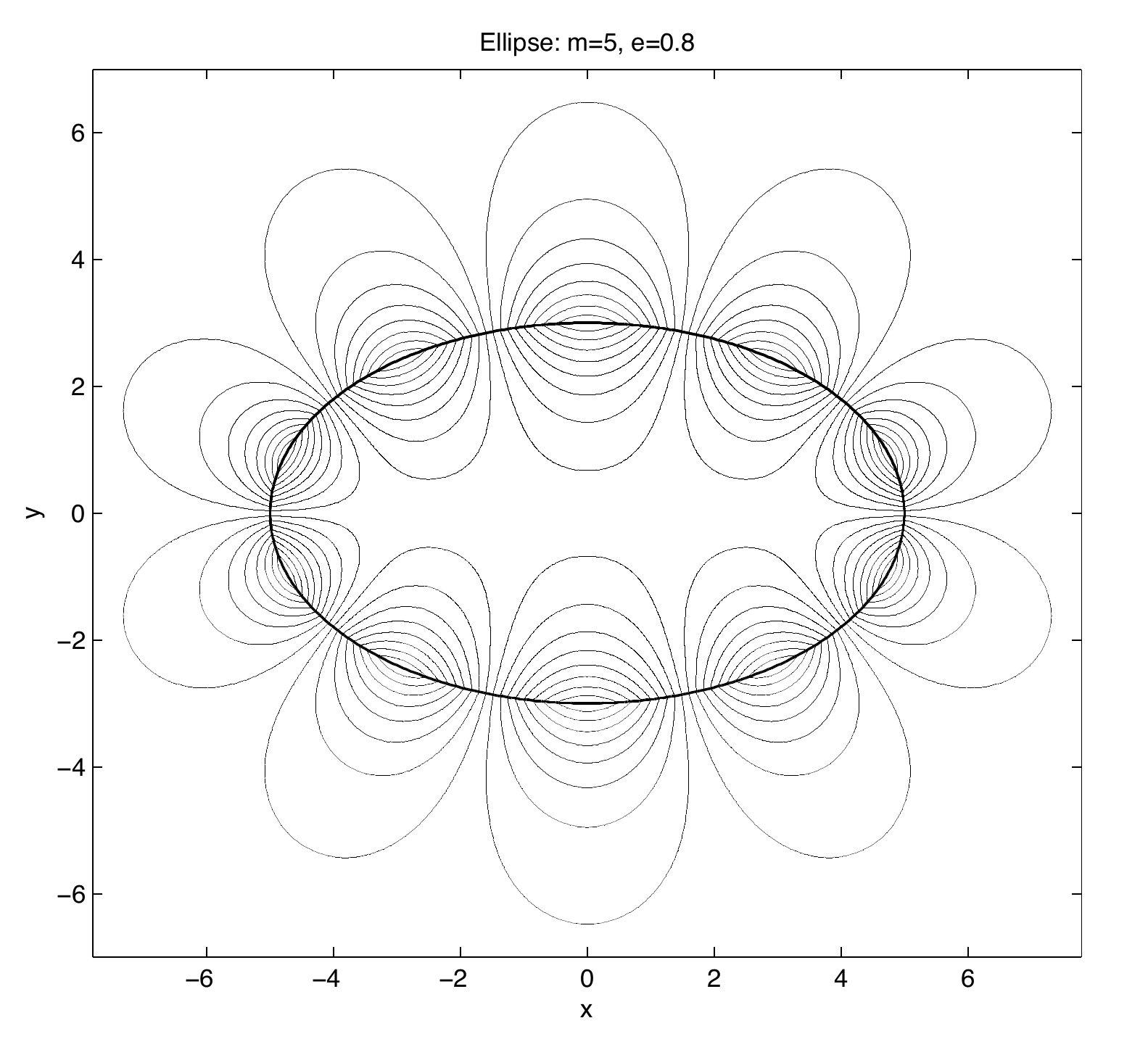}
\caption{\label{elliptical5} Typical streamlines corresponding to multipolar vortices around an ellipse, with $m=5$. The harmonicity of the associated velocity field is given by $\sin m v$.}
\end{figure}

\begin{figure}[!htb]
\centering
\includegraphics[width=9cm]{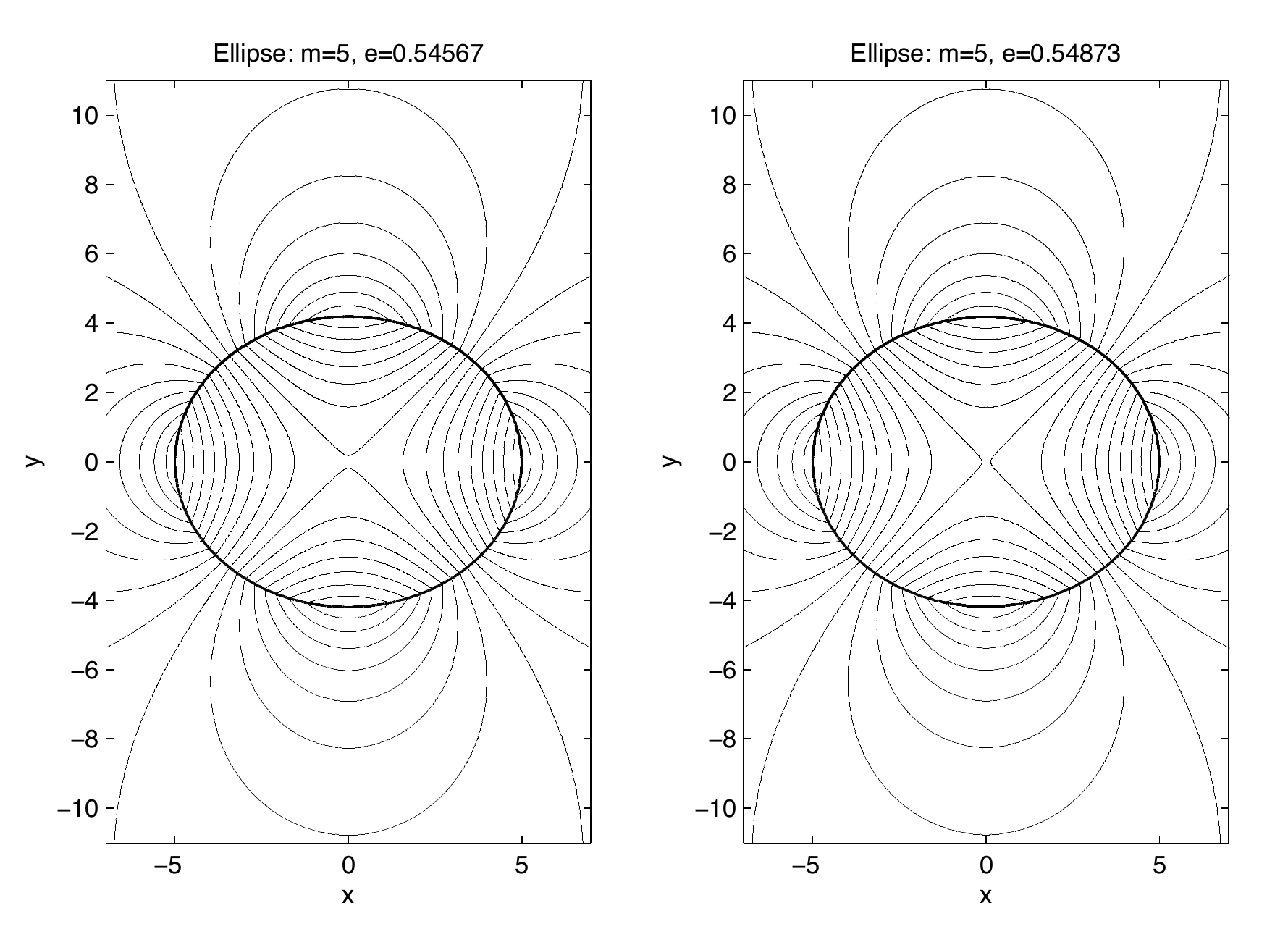}
\caption{\label{twoellipses2} Typical streamlines corresponding to multipolar vortices around two ellipses, with $m=2$, but different eccentricity $e$. The harmonicity of the associated velocity fields is given by $\cos m v$.}
\end{figure}

Figure~\ref{polar5} shows the distribution of streamlines inside and outside a circle of radius $r_0=5$, with $m=5$, and with harmonicity $\cos m \varphi$. The resulting street of 10 vortices equally spaced around the perimeter of the circle may be compared with Figure 8 in Crowdy's paper~\cite{Crowdy:2004}. Figure~\ref{elliptical5} shows a similar configuration of vortices, still for $m=5$, but now distributed along the perimeter of an ellipse, and with harmonicity different from the previous plot, $\sin m \varphi$. An immediately observable difference between this Figure and Figure~\ref{polar5} is that the vortices are no longer all of the same size. The vortices nearer the origin are larger than those farther from the origin, a behavior caused by the particular scaling factors associated with elliptic coordinates. Changing the angular dependence between a cosine function and a sine function simply results, in these cases, in a $\pi/2m$-shift of the angular variable. Note, for example, that in Figure~\ref{polar5} there is a separatrix along the $y$ axis, while in Figure~\ref{elliptical5}, the separatrix coincides with the $x$ axis. From the previous figures we may observe that as the eccentricity of an ellipse increases, the vortices on opposing sides of the major axis grow closer together. Eventually, as may be seen in Figure~\ref{twoellipses2}, which shows streamlines inside and outside an ellipse, with $m=2$, two streamlines may join, giving rise to distinctive cat's-eye patterns, similar to those typically associated with Stuart vortices~\cite{Stuart:1967}. In the plot on the left, the two streamlines are not yet joined, while on the plot on the right, around an ellipse with greater eccentricity, the two streamlines have coalesced. All of the above ellipses have been plotted in such a way that the length of the major axes is equal to 10, and the eccentricity $e$ is then varied by giving different values to the focal distance $f$.

\begin{figure}[!htb]
\centering
\includegraphics[width=9cm]{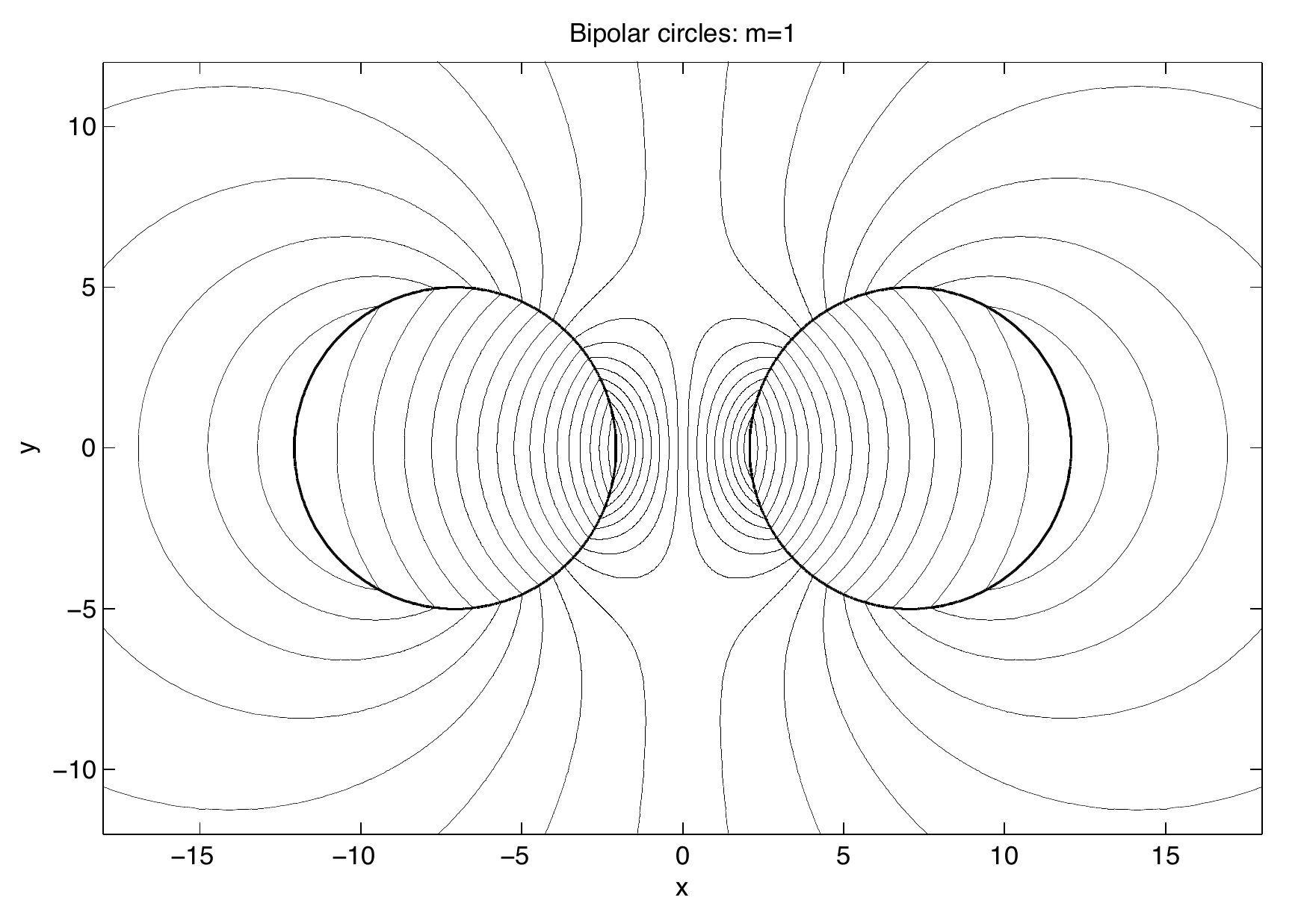}
\caption{\label{bipolar1cos} Typical streamlines corresponding to multipolar vortices around bipolar circles, with $m=1$. The harmonicity of the associated velocity field is given by $\cos m \sigma$.}
\end{figure}

\begin{figure}[!htb]
\centering
\includegraphics[width=9cm]{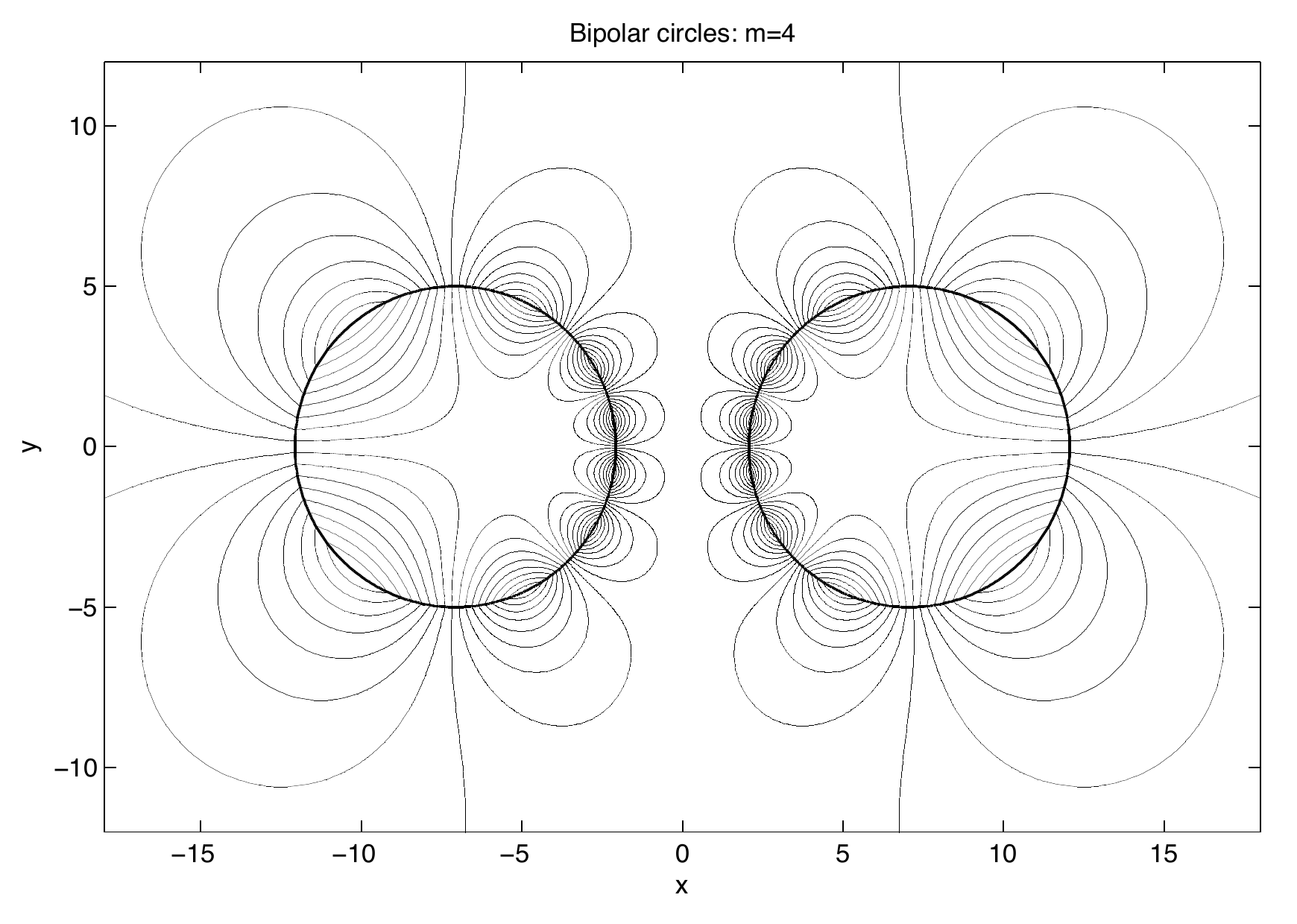}
\caption{\label{bipolar4sin} Typical streamlines corresponding to multipolar vortices around bipolar circles, with $m=4$. The harmonicity of the associated velocity field is given by $\sin m \sigma$.}
\end{figure}

Figures~\ref{bipolar1cos} and \ref{bipolar4sin} show typical streamline distributions corresponding to bipolar coordinates with $m=1$ and $m=4$, respectively. In both cases we have set $r_0=5$ and $a=5$, so that $\tau_0 = a \, \mbox{csch } (r_0/a) = 4.2546$, as we see from the Appendix. For $m=1$ we observe two dipoles, distributed symmetrically on opposite sides of the $y$ axis, and for $m=4$ we observe two octupoles, which are also distributed symmetrically on opposite sides of the $y$ axis, as before, but with the $x$ axis now acting as a separatrix as well, due to the different harmonicity of the angular dependence. A distinctive feature of the vortical structures which form around bipolar circles, again due to the presence of scaling factors, is that the relative size of the structures can vary significantly -- vortices near the origin are noticeably smaller than those farther away from the origin.  

\section{Line vortices and patch vortices}
\label{linevortices}

Point vortices and uniform vortex patches are the most widely studied vortex structures, and they are closely related. The radius of a Rankine vortex, which is a classical example of a vortex patch, can be made to tend to zero while its vorticity tends to infinity in such a way that its circulation is constant. The structure arrived at in this manner is a point vortex. Conversely, a point-vortex solution may be desingularized by smearing out the vorticity to a uniform patch of nonzero area, as is done, for example, by Dritschel~\cite{Dritschel:1985}. Both types of vortices have been used to model multipolar vortex equilibria. Crowdy and Marshall~\cite{Crowdy:2004hh} use a finite distribution of point vortices superimposed on a uniform vortex patch to model multipolar vortices, while Aref and Vainchtein~\cite{Aref:1998} construct complicated point-vortex equilibria by considering a frame of reference corotating with one such equilibrium, and ''growing'' point vortices at well-chosen corotating points. 

The discussion in this section deals with connections of the three complete and orthogonal sets of harmonic vortices introduced in this article with these other types of vortices on a plane, but considering line vortices instead of point vortices, as required by our formulation. We use the Green's function for Poisson's equation in standard fashion to obtain representations of both patch vortices and line vortices in terms of the sets of harmonic vortices given in section~\ref{harmonicvortices}, and we then construct Rankine vortices for different geometries using these representations.  

\subsection{The Green's function for Poisson's equation}

The Green's function $G$ for Poisson's equation, corresponding to unit line vortex sources, has the following forms and expansions in polar, elliptic and bipolar harmonics, respectively:

\begin{align}
G(r,\varphi;r',\varphi') &= -\frac{1}{2\pi} \ln | \vec{r} - \vec{r'} | = -\frac{1}{2\pi} \ln r_> + \frac{1}{\pi} \sum\limits_{m=1}^{\infty} \frac{r_<^m}{r_>^m} \frac{\cos m(\varphi - \varphi')}{m},
\label{polargreen} \\
G(u,v;u',v') &= -\frac{1}{2 \pi} u_> + \frac{1}{\pi} \sum\limits_{m=1}^{\infty} \frac{1}{m} \left( \cosh mu_< \, e^{-mu_>} \, \cos mv \, \cos mv' \right.
\notag \\
&\left. + \sinh mu_< \, e^{-mu_>} \, \sin mv \, \sin mv' \right), 
\label{ellipticalgreen} \\
G(\tau,\sigma;\tau',\sigma') &= \frac{1}{2\pi} \left( \ln | \vec{r} + \vec{r'} | - \ln | \vec{r} - \vec{r'} | \right) = \frac{1}{2\pi} \left( \tau_< - \tau_> \right) 
\notag \\
&+ \frac{1}{\pi} \sum\limits_{m=1}^{\infty} \frac{1}{m} \, e^{m \tau_<} \, \sinh m \tau_> \left( \cos m \sigma \, \cos m \sigma' + \sin m \sigma \, \sin m \sigma' \right)
\notag \\
&+ \frac{1}{\pi} \sum\limits_{m=1}^{\infty} \frac{1}{m} \, e^{-m \tau_>} \, \sinh m \tau_< \left( \cos m \sigma \, \cos m \sigma' + \sin m \sigma \, \sin m \sigma' \right),
\label{bipolargreen}  
\end{align}

\noindent
where $\vec{r}$ and $\vec{r'}$ are position vectors of the field and source points. Their respective harmonic compositions are easily appreciated in the above expressions, and, further, a literal reading of each term suggests the distinction between the harmonic functions to be used inside and outside the coordinate boundary circles or ellipses along which the vorticity is distributed.

For any vorticity field $\vec{\omega} ( \vec{r} ) = \hat{k} \omega ( \vec{r} )$ distributed on the plane, the corresponding vector potential is given by the following integral, with the corresponding Green's function as a kernel:

\begin{equation}
\vec{A} (\vec{r}) = \hat{k} A(\vec{r}) =
\hat{k} \int h_1 (q_1',q_2') \, h_2 (q_1',q_2') \, G(q_1,q_2;q_1',q_2') \, \omega (q_1',q_2') \, dq_1' \, dq_2'.
\label{vectorfield}
\end{equation}

\noindent
As an illustration, in the following section we use this equation to construct the familiar Rankine vortex.

\subsection{Rankine vortices}

The Rankine vortex, as defined by Saffman~\cite{Saffman:1992} and Crowdy~\cite{Crowdy:1999,Crowdy:2002a}, is a uniform, rectilinear, circular vortex filament, with constant vorticity $\omega_0$ distributed inside a circle of radius $r=r_0$. Specifically, it is given by

\begin{equation}
u_{\varphi} = \left\{
    \begin{array}{l@{\quad}l}
      \dfrac{\omega_0 \,r}{2} &  r < r_0 \\
      \dfrac{\omega_0 \, r_0^2}{2r} & r > r_0
    \end{array}
  \right. , \\
\label{rankine}
\end{equation}

\noindent
where $u_{\varphi}$ is the angular component of the velocity $\vec{u}$. We now use equation (\ref{vectorfield}) to calculate the appropriate vector potential inside the circle. Observe that in this case, calculation of the integral in (\ref{vectorfield}) in polar coordinates involves the surface element $r \, dr \, d \varphi$, so that only the $m=0$ term in equation (\ref{polargreen}) contributes, as the higher harmonic angular contributions all vanish. If we now write the vorticity simply as $\omega = \hat{k} \omega_0$, the vector potential inside the source circle, with radius $r=r_0$, is

\begin{equation}
\vec{A}^i = - \hat{k} \, \frac{\omega_0}{4} \left[ r^2 + r_0^2 (2 \, \ln \, r_0 -1 ) \right].
\label{potentialrankine}
\end{equation}

\noindent
This potential leads to the same velocity field inside the circle given in (\ref{rankine}). The Rankine vortex is completed by using the same form of the vector potential given in equation (\ref{polar0}) outside the source circle, with $A_0=-(\omega_0 \, r_0^2)/2$. The respective streamlines are represented by concentric circular lines.

Taking into account the above representation of the circular Rankine vortex in terms of circular harmonics, we may propose an extension of the Rankine vortex in both elliptic and bipolar geometries. The elliptic and bipolar vorticity distributions for the respective Rankine-like vortices inside the respective boundaries are chosen as follows:

\begin{align}
\vec{\omega} (u,v) &= \hat{k} \frac{\Gamma}{h_u \, h_v},
\label{ellipticalrankine} \\
\vec{\omega} (\tau, \sigma) &= \mp \hat{k} \frac{\Gamma}{h_{\tau} \, h_{\sigma}} \quad \mbox{for } \tau \lessgtr 0,
\label{bipolarrankine}
\end{align}

\noindent
where $\Gamma$ is the strength of the vortex filaments. The scale factors in equation (\ref{vectorfield}) and the above expressions for the vorticities allow for the direct integration of the vector potentials inside the boundaries, and the calculation of the corresponding velocity fields thereof. For the elliptic Rankine vortex we obtain

\begin{equation}
\vec{u}^i = \hat{v} \dfrac{\Gamma \, u}{h_u},
\label{ellipticalrankinevel}
\end{equation}

\noindent
and for the bipolar Rankine vortex we obtain

\begin{equation}
\vec{u}^i = \hat{\sigma} \dfrac{\Gamma \, \tau}{h_{\tau}}.
\label{bipolarrankinevel}
\end{equation}

\noindent
The vorticities given in (\ref{ellipticalrankine}) and (\ref{bipolarrankine}) may then be recovered from the corresponding velocity fields above, as expected. Outside the ellipse and the circles, the velocity varies in the same manner as in (\ref{ellipticalvelocitye0}) and (\ref{bipolarvelocitye0}), respectively, so that the vorticity outside the boundaries is zero in both cases. The elliptic Rankine vortex is represented by confocal elliptic streamlines, and the bipolar Rankine vortex is represented by pairs of nested circular streamlines.

The shielded Rankine vortices proposed by Crowdy~\cite{Crowdy:2004hh} are constructed as the superposition of a Rankine vortex and a line vortex at its center with a circulation of the same magnitude but directed in the opposite sense, thus ensuring that the velocity field vanishes for $r > r_0$, and also that the combined structure has zero total circulation. The angular component of the velocity is now written as

\begin{equation}
u_{\varphi} = \left\{
    \begin{array}{l@{\quad}l}
      \dfrac{\omega_0 \,r}{2} - \dfrac{\omega_0 \, r_0^2}{2r} &  r < r_0 \\
      0 & r > r_0
    \end{array}
  \right. . \\
\label{shieldedrankine}
\end{equation}

\noindent
In our formulation, the line vortex at the origin of circular coordinates can be seen as the limiting situation in which $r_0 \to 0$: the velocity field inside does not appear, and the velocity field outside is given by equation (\ref{polarvelocitye0}). This line vortex can then be superposed on the circular Rankine vortex obtained previously to produce a shielded Rankine vortex. This is essentially the same treatment as that proposed by Morel and Carton~\cite{Morel:1994}, who express the shielded Rankine vortex as a limit of the two-contour Rankine vortex. Alternatively, we may arrive at the shielded Rankine vortex by superposing the vector potential (\ref{polar0}) outside the source circle onto the existing Rankine vortex, with $A_0$ as before, but with opposite sign. In this manner, the shielded Rankine vortex is obtained without placing a line vortex within the existing patch vortex, and as a result the velocity field inside the circular boundary is everywhere non-singular.

As natural geometric extensions of the circular shielded Rankine vortex, we consider shielded Rankine-like vortices in elliptic and bipolar coordinates. For the first case, as $u_0 \to 0$, the ellipse, with $e \to 1$, takes on the form of two straight line segments, one just above the $x$ axis, the other just below, going around the two foci. The velocity field outside this degenerate ellipse is given by equation (\ref{ellipticalvelocitye0}), and the corresponding streamlines will be confocal coordinate ellipses. In bipolar coordinates, as $\tau \to \pm \infty$ we obtain two line vortices of opposite polarities, centered at $(x,y) = (\pm a, 0)$. In this case, the velocity field outside would be given by equation (\ref{bipolarvelocitye0}), corresponding to the bipolar vortex with $m=0$, and with streamlines in the form of nested coordinate circles which cover the entire plane. The shielded form of the corresponding vortices may be arrived at in the same manner as that described above for the case of circular vortices. 

\section{Discussion and connections with other types of vortices}
\label{discussion} 

Vortices with bipolar and elliptic geometries have been the subject of previous studies, and their physical feasibility has been well established, albeit for the case of rotating vortices. Kirchhoff~\cite{Kirchhoff:1876} generalized the idea of a Rankine vortex by proving that isolated two-dimensional vortex patch ellipses are exact solutions of the nonlinear Euler equations. The associated flow is nonsteady, with the elliptic patch rotating steadily about its center. The stability of these patch vortices has been shown by Mitchell and Rossi~\cite{Mitchell:2008} to depend precisely on the eccentricity of the ellipse - we have shown in section~\ref{graphical} that the eccentricity of the boundary ellipse does indeed affect the behavior of the associated streamlines. Moore and Saffman~\cite{Moore:1971} have, in turn, generalized the Kirchhoff elliptic vortex to an elliptic vortex patch in a uniform straining field.

With regard to dipolar vortices, the best-known example is the Lamb--Chaplygin dipole, with Chaplygin independently deriving the same dipole solution as that outlined by Lamb in his famous book~\cite{Lamb:1895}. Chaplygin's work on two-dimensional coherent vortex structures in an inviscid fluid was originally published over a hundred years ago~\cite{Chaplygin:1903} and was recently translated and commented by Meleshko and Van Heijst~\cite{Chaplygin:2007,Meleshko:1994}. The Lamb--Chaplygin dipole consists of a dipolar vortex structure with continuous vorticity distributed inside a circle. Our own comments on the original description of the dipole inside a circle have to do with recognizing the possibility of extending this description to higher harmonic vortices. We observe that the equation for the vector potential is simply Poisson's equation with a source term chosen to be proportional to the vector potential itself. When the latter is chosen specifically as the product between a radial function and an angular function of the form $\sin \varphi$, the differential equation for the radial function turns out to be Bessel's equation for $m=1$. By replacing $\sin \varphi$ by $\sin m \varphi$ or $\cos m \varphi$, with $m=2,3,4,\ldots$, we obtain higher harmonicity vorticity sources, accompanied by the corresponding radial Bessel functions of order $m$.

Extensions of vortices on a plane for higher harmonics and for other geometries have been the subject of investigations by Crowdy, who studies Stuart vortices on a plane~\cite{Crowdy:2003ef} and a sphere~\cite{Crowdy:2004}, and by Bogomolov~\cite{Bogomolov:1979} and Kimura and Okamoto~\cite{Kimura:1987} on a sphere, amongst others. These papers suggest that stereographic projections from the plane onto spherical surfaces can be performed on the complete sets of vortices on a plane presented in this paper, allowing for comparisons with the corresponding harmonic vortices on such surfaces. 

On the other hand, we point out that our approach may also be used to identify complete sets of spherical, spheroconal, and prolate and oblate spheroidal harmonic vortices in three dimensions, inside and outside spherical or spheroidal boundary coordinate surfaces, where the vorticity fields are distributed.

The solutions examined in this paper represent coherent vortical structures characterized by a continuous distribution of nonvanishing vorticity on circular and elliptic boundaries. The associated streamlines differ from those normally obtained in the context of multipolar vortex structures (see Crowdy~\cite{Crowdy:1999}) in that they are not confined to a bounded region of the plane, but rather, they cover the whole plane, inside and outside the corresponding boundaries. We do observe separatrix streamlines, and their number and geometrical distribution depend critically on the harmonicity of the vorticity sources. Although these solutions are not, in general, consistent with observed multipolar vortical structures, they do share several properties in common with such structures, and it must be stressed that the completeness of our set of solutions allows for the expression of well-known solutions to the two-dimensional steady Euler equations in terms of series expansions of harmonic functions. As an example, we have written both the circular Rankine vortex and the circular shielded Rankine vortex in terms of the harmonic solutions obtained above, and we have extended the idea of Rankine vortices to elliptic and bipolar geometries. These structures may be further extended to higher harmonicities, thus, in principle, allowing for the description of multipolar equilibria of the various forms discussed in section~\ref{intro}. These extensions to other types of vortices, both on the plane and on the surface of a sphere, will be presented in future works. 

\appendix

\section{Polar, elliptic, and bipolar coordinates}

\begin{enumerate}

\item
{\bf Polar coordinates}

The relationship between the polar coordinates $(r, \varphi)$ and Cartesian coordinates $(x,y)$ is expressed as follows:

\begin{eqnarray*}
x & = & r \cos \varphi, \\
y & = & r \sin \varphi,
\end{eqnarray*}

\noindent
where $0 \leqslant r < \infty$ and $0\leqslant \varphi \leqslant 2\pi$. Conversely, polar coordinates are expressed in terms of Cartesian coordinates through

\[
r^2=x^2+y^2,
\]

\noindent
which represents a family of circles with radius $r$ centered on the origin, and 

\[
\varphi = \tan ^{-1}\left( y/x \right),
\]

\noindent
a family of straight lines which pass through the origin with slope $\tan \varphi$.

\item
{\bf Elliptic coordinates}

Elliptic coordinates $(u,v)$ are related to Cartesian coordinates $(x,y)$ by means of the following transformation:

\begin{eqnarray*}
x & = & f\cosh u \cos v, \\
y & = & f\sinh u \sin v,
\end{eqnarray*}

\noindent
where $0 \leqslant u < \infty$ and $0 \leqslant v \leqslant 2 \pi$. In this case, the inversion is given by

\[
\frac{x^2}{f^2 \cosh ^2 u}+\frac{y^2}{f^2 \sinh ^2 u}=1,
\]

\noindent
which represent ellipses centered at the origin, foci at $( x = \pm f, y = 0 )$, major semiaxes  $f \cosh u$, minor semiaxes $f \sinh u$, and eccentricity $e = 1 / \cosh u$, and

\[
\frac{x^2}{f^2 \cos ^2 v}-\frac{y^2}{f^2 \sin ^2 v}=1,
\]

\noindent
which are hyperbolas which share foci with the above family of ellipses, with real semiaxes $f \cos v$, imaginary semiaxes $f \sin v$, and eccentricity $e = 1 / \cos v$.

\item
{\bf Bipolar coordinates}  

Bipolar coordinates $(\sigma, \tau)$ are defined in terms of Cartesian coordinates $(x,y)$ as follows:

\begin{eqnarray*}
x & = & \frac{a \sinh \tau }{\cosh \tau -\cos \sigma }, \\
y & = & \frac{a \sin \sigma }{\cosh \tau -\cos \sigma },
\end{eqnarray*}

\noindent
where $( -\infty <\tau <\infty , 0 \leqslant \sigma \leqslant 2\pi )$. The inverse transformation is given by

\[
\left( x - a \coth \tau \right)^2 + y^2 = a^2 \mbox{csch}^2 \tau,
\]

\noindent
and

\[
x^2 +\left( y - a \cot \sigma \right)^2 = a^2 \csc^2 \sigma.
\]

\noindent
The first of these expressions describes a family of nested circles with radii $a\, \mbox{csch} \tau $ centered at $( x = a \, \coth \tau , y=0 )$. For each fixed value of $|\tau|$, we obtain a couple of nonintersecting circles placed symmetrically on either side of the $y$ axis. The second expression represents a family of circles with radii $a \, \mbox{csc}\sigma $ centered at $( x = 0, y = a \cot \sigma)$, where circles are distributed symmetrically on either side of the $x$ axis, with intersection points at $( x = \pm a, y=0 )$.

\end{enumerate}

\subsection*{Acknowledgements}

The authors wish to thank Christian Esparza L\'opez for his helpful comments. The authors gratefully acknowledge financial support for this research from DGAPA UNAM through project PAPIIT IN109214, and from CONACYT through SNI 1796.

\bibliographystyle{sigma}
\bibliography{vortex}

\LastPageEnding

\end{document}